\documentclass{article}
\usepackage{graphicx}
\usepackage{enumerate}
\usepackage[square]{natbib}
\usepackage{authblk} 
\usepackage{amsthm} 
\usepackage{amsfonts}
\usepackage{physics} 
\usepackage{paralist} 
\usepackage{multirow} 

\newtheorem{mydef}{Definition} 

\date{\today}

\begin{document}

\title{The conception of reality in Quantum Mechanics}
\author[1]{Gerd Ch.Krizek}
\affil{University of Vienna - Quantum particle group}
\affil{UAS Technikum Wien - Department of Applied Mathematics and Sciences}


\maketitle

\begin{abstract}
\noindent 
Disputes on the foundations of Quantum Mechanics often involve the conception of reality, without a clear definition on which aspect of this broad concept of reality one refers. We provide an overview of conceptions of reality in classical physics, in philosophy of science and in Quantum Mechanics and its interpretations and analyse their differences and subtleties. Structural realism as conception in philosophy of science will turn out to be a promising candidate to settle old problems regarding the conception of reality in Quantum Mechanics. We amend the analysis of three main interpretations and their relation to structural realism by three well-known informational approaches in the interpretations of Quantum Mechanics. Last we propose an additional class of structural realism supported by QBism. 

\end{abstract}



\newpage

\section{Introduction}
\parindent 0cm       

Quantum Mechanics is one of the most successful physical theories; still, its interpretation and implications are a matter of dispute. One conception that is at the very heart of the debates is the conception of reality. Realism is a broad conception in philosophy of science; therefore, the term reality is inherently ambiguous. \newline

Reality represents a variety of conceptions\cite[]{sankey2008scientific}. Disputes on the foundations of Quantum Mechanics often involve the conception of reality, without a clear definition on which aspect of this broad conception of reality one refers. The scope of this work is to give an overview on the conceptions of reality present in the EPR argument, Bells inequality and the interpretations of Quantum Mechanics. \newline

The development of the theory of relativity was accompanied by the advent of positivistic philosophy, which strongly influenced the development of Quantum Mechanics. Since the theory of relativity strongly followed Empiricism, which asserts that only what is measurable can be seen as real, this stance was adopted by most of the founding fathers of Quantum Mechanics. Not all of them welcomed the anti-realist philosophy that pervaded the early years of Quantum Mechanics. Einstein and Schr\"odinger were strong opponents on the realist-side to Bohr, Heisenberg and Pauli, who advanced the anti-realist position. An overview on this famous debate is provided by \cite[]{landsman2006champions}. 

What the Einstein-Bohr debate initialised was an ongoing struggle on the very nature of reality. After a brief outline of the realist accounts in classical physics and philosophy of science we present the definitions of this disputed realities and their involvement in the EPR argument, the Bell inequality and the measurement process.\newline

We amend a the presentation of \cite[]{esfeld2013ontic} on the relation of structural realism and Quantum Mechanics, by three more interpretations of Quantum Mechanics that involve the concept of information and provide candidates for the structures proposed by structural realism. 

\newpage

\section{Reality in classical physics}

Reality is a broad conception, so we want to provide an overview on its aspects as well as its presumptions and clarify its definitions for the conceptual use of reality. Since the conception of reality will change vastly in Quantum Mechanics, first we want to provide an overview on aspects of classical realism, which are relevant to our discussion on reality in Quantum Mechanics. As \cite[p.14]{braver2007thing} points out realism in not always defined in a straightforward way:

\begin{quote}
\textit{``Of course, there are as many species of realism as there are kinds of objects, since realism can be type-specific or "local" rather than "global": one can be a realist about stars but not about the occurrence of the sequence "7777" in pi; a realist about the past, but not about the future; a realist about unobserved, but not unobservables; or the reverse for each pair.''}
\end{quote}

One could have the impression that "reality" is a far too broad conception to give a clear definition. \cite[]{gallie1955essentially} provided a framework for terms that represent such a broad conception, the idea of Essentially Contested Concepts:

\begin{quote}
\textit{``I shall try to show that there are disputes, centred on the concepts which I have just mentioned, which are perfectly genuine: which, although not resolvable by argument of any kind, are nevertheless sustained by perfectly respectable arguments and evidence. This is what I mean by saying that there are concepts which are essentially contested, concepts the proper use of which inevitably involves endless disputes about their proper uses on the part of their users.''}
\end{quote}

Essentially Contested Concepts usually refer to sociopolitical terms such as "democracy" and "freedom," but the term "reality" with its variety of interpretations gives the impression of being a good candidate for an Essentially Contested Concept. We will not examine in detail if "reality" conforms to the conditions of Essentially Contested Concepts, but want to adopt the spirit of the discussion on Essentially Contested Concepts for our examination of the term reality. 
\newpage

\subsection{The structure of a physical theory and its metaphysics}

We proposed a classification scheme to illustrate the formal, conceptual and ontological differences between various interpretations of Quantum Mechanics \cite[]{krizek2017classification}. A visualization of the scheme shows the assignment of the different elements of a theory and the related metaphysics. 


\begin{table}[h]
\centering

\label{Tab_scheme}
\begin{tabular}{|l|l|c|c|}
\hline
1 & \begin{tabular}[c]{@{}c@{}}\\ \includegraphics[width=0.1\textwidth, height=30px]{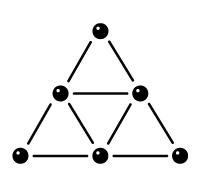} \\ \\ \end{tabular} & \begin{tabular}[c]{@{}c@{}}\\ Mathematical laws \\ between mathematical symbols \\ \\ \end{tabular} & \multirow{3}{*}{Physics} \\ \cline{1-3}
2 & \begin{tabular}[c]{@{}c@{}}\\ \includegraphics[width=0.1\textwidth, height=30px]{1} \\ \\ \end{tabular} & \begin{tabular}[c]{@{}c@{}}  \\ Interpretation physical quantities \\ Relation physical quantities - Measurements \\ Measurement laws \\ \\ \end{tabular} &                          \\ \cline{1-3}
3 & \begin{tabular}[c]{@{}c@{}}\\ \includegraphics[width=0.1\textwidth, height=30px]{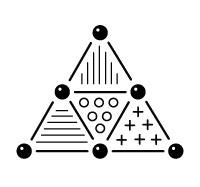} \\ \\ \end{tabular} & \begin{tabular}[c]{@{}c@{}} \\ Theoretical terms  \\Concepts and principles  \\\\ \end{tabular} &                          \\ \hline
4 & \begin{tabular}[c]{@{}c@{}}\\ \includegraphics[width=0.1\textwidth, height=40px]{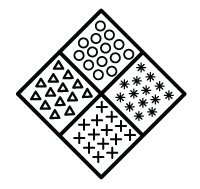} \\ \\ \end{tabular} & \begin{tabular}[c]{@{}c@{}}\\ Ontology - beables - \\ elements of reality \\ \\ \end{tabular} & Metaphysics               \\ \hline
\end{tabular}
\caption{The Classification scheme and its application}
\end{table}

On the mathematical level only mathematical symbols and their interrelation are introduced. 

On level two of the classification scheme, names f or the
symbolically represented quantities are assigned. This is the introduction of the
basis of a theoretical term. The full framework of theoretical terms is given by
conceptual level three of the classification scheme. It contains principles and conceptions. Theoretical terms are fully defined on that level. An overview on the philosophical problems of theoretical terms is provided by \cite[]{sep-theoretical-terms-science}.\newline

In the framework of the evolution of physical theories, the question on the status of theoretical terms arises again. Theoretical terms in several theories might be represented by the same semantical term, but the meaning might differ vastly as  emphasized by \cite[p.30]{esfeld2008naturphilosophie}: 

\begin{quote}
\textit{``Der Wechsel von einer alten zu einer neuen Theorie kann derart schwerwiegend sein, dass die inferentiellen Beziehungen, die den Inhalt der charakteristischen Begriffe der beiden Theorien definieren, sich so radikal \"andern, dass die charakteristischen Begriffe der alten Theorie nicht in den charakteristischen Begriffen der neuen Theorie formuliert werden k\"onnen. Es gibt kein gemeinsames Ma\ss \ f\"ur die charakteristischen Begriffe beider Theorien.''}
\end{quote}

\begin{quote}
\textit{``The change from an old to a new theory can be radical in that sense, that the inferential relations, which define the content of the characteristic terms of booth theories, change radically in a way that the characteristic terms of the old theory cannot be formulated in words of the characteristic terms of the new theory. There is no common measure for the characteristic terms of both theories.\footnote{Translation by the author.}''}
\end{quote}

This concept is called Incommensurability and has been introduced by \cite[]{kuhn1962structure} and \cite[]{feyerabend1993against}. In Quantum Mechanics the term "particle" does not have the same conceptual and ontological meaning as in classical particle mechanics. \newline

Detailed accounts on the historical debates on the structure of scientific theories are provided by \cite[]{suppe1977structure}, \cite[]{sneed1976philosophical}, \cite{ludwig1978grundstrukturen} and \cite[]{ludwig2007new}. \newline

Following this brief overview on the structure of physical theories we provide a list of conceptions that contribute to the conception of classical realism. 

\subsection{Uniqueness}

In classical physics we refer to old principles that build the very basis of scientific endeavour. This basis has been set by the Greek philosophers, especially Aristotle, who provided a logical system and a philosophy of nature. \newline

Two of these basic principles influence our scientific thinking in a fundamental way: The contradiction principle and the Law of the Excluded Middle. 

The contradiction principle is formulated in the following way \cite[]{ross1928works}:
\begin{quote}
\textit{``It is, that the same attribute cannot at the same time belong and not belong to the same subject and in the same respect.''}
\end{quote}

The contradiction principle can be seen as logical rule, as epistemological, or ontological rule. In that sense, it refers to different regimes of scientific thinking. \newline

Depending on the method of interpreting Aristotles contradiction principle, it is related to a different level of the classification scheme we presented in \cite[]{krizek2017classification}. The scheme consists of four levels that represent the structure of a scientific theory: The mathematical formalism, the assignment to measurement values and theory-terms, the conceptual structure, and the ontology of the theory. \newline

The logical contradiction principle would refer to the mathematical formalism on level one, the epistemological\footnote{Since the epistomological application is the one which is not straightforward, we want to give an example: In a political system like democracy, people are seen equal by principle. All citizens are allowed to elect. All non-citizens are not allowed to elect. Though we know that the individuals are not equal on an ontological level, in the conception of democracy an epistemological equality is adopted, which gives no implication to the ontological level for the individual human.} contradiction principle would refer to the level three of conceptions and principles of a theory, and the ontological contradiction principle would refer to the level four of the classification scheme. \newline

In a formal way the contradiction principle is given by the statement on any statement $p$: 

\begin{equation}
 \lnot(p \wedge \lnot p)
\end{equation}

The law of the excluded middle states that between a statement $p$ and its negation $\lnot p$ there is nothing in between. Therefore, in a formal way, the law of the excluded middle is given by \newline

\begin{equation}
p \vee \lnot p 
\end{equation}

These logical principles follow a priori from our common sense experience that objects are here or there, from the behaviour we observe in nature. This a priori knowledge influenced our thinking and the scientific method tremendously. So the state of existence of an ontological object is thought as unique, as unambiguous. This uniqueness is one essential basis of classical reality. \newline 

Quantum Mechanics with superposition, entanglement and non-classical statistics challenged this view on the ontological level. Some interpretations, such as the Quantum logic approache challenged also the mathematical formalism and questioned the classical logical structure \cite[]{mittelstaedt1978quantum}.  

\subsection{Independence}

In classical physics, the conception of reality is independent of us, the human beings as observers. Reality was and will be there, independent if we describe it or not \cite[p.15ff]{sankey2008scientific}. And the term "description" gives a good picture of the idea behind independence in this context. A description is the process of observation without altering the observed. It describes some reality that is there in a language, or another formal way. This independence will be realised within the definition of metaphysical realism in philosophy of science. 

\subsection{Completeness and correspondence}

Directly related to the notion of independence is the conception of correspondence and completeness. The idea of correspondence is that the scientific theories and their content corresponds to something out there. We observe phenomena and describe them with the formal language of mathematics and a semantic structure of terms and statements. It is one assumption of realism that this formal and semantic structure corresponds to a reality that is the causation of the observed phenomena \cite[]{putnam1981reason} and \cite[]{putnam1988representation}. \newline

The correspondence between reality and the formal and semantic description of the theory can involve different levels of the theory structure. We allude to the classification scheme and can distinguish pure mathematical correspondence and a conceptual correspondence. In a mathematical correspondence, the elements of the mathematical formalism of the theory correspond to elements on the ontological level. In the conceptual correspondence, concepts are corresponding to elements on the ontological level. \newline

We provided a definition of different conceptions of completeness in \cite[p.26ff]{krizek2017einstein}. We summarize these conceptions of completeness and their relation to reality: 

\paragraph{Theory completeness:}

\begin{mydef}
\label{Definition theory completeness}
A theory fulfills theory completeness if its physical quantities and laws are connected by concepts that only relate only to each other and the theory renounces extra assumptions of concepts, which exceed the framework of the theory. 
\end{mydef}

\paragraph{Bijective completeness:}

\begin{mydef}
\label{Definition bijective completeness}
Bijective completeness means that every element of the physical reality $\epsilon_i $ must have a counterpart  $ t_i $ in the physical theory.
\end{mydef}

\paragraph{Born completeness:}

\begin{mydef}
\label{Definition Born completeness}
Born completeness for a theory is fulfilled if the theory contains no probabilistic elements. A theory which contains statistical accounts and probabilistic statements is considered to be incomplete. 
\end{mydef}

\newpage

\paragraph{Schr\"odinger completeness:}

\begin{mydef}
\label{Definition Schrodinger completeness}
Schr\"odinger completeness for a theory is fulfilled if the theory contains probabilistic statements about elements of the theory. The probabilistic statements refer not to the statistical ensemble of many events, but to one single event. The probabilistic statements are fundamental in the sense that nothing more can be said about the system. 
The theory is considered to be incomplete if there are potential new elements\footnote{Einstein calls them "fremde Faktoren"(external factors). The translation of "fremde" is peculiar, since in the literally translation it would be "strange," "weird," "alien" or "different," but in context it refers to factors that are not considered in the theory. They are not part of the theory, so we decided to use external. In our modern context, these factors could be understood as "hidden variables."} in the theory that would abandon the probabilistic character of the theory. 
\end{mydef}

\paragraph{$\psi$ - completeness:}

\begin{quote}
\textit{``An ontological model is $\psi$-complete if the ontic state space $\Lambda$ is isomorphic to the projective Hilbert space $\mathcal{PH}$ (the space of rays of Hilbert space) and if every preparation  procedure $P_\psi $ associated  in  quantum theory with a given ray $\psi$ is associated in the ontological model with a Dirac delta function centered at the ontic state $\Lambda$ ψ that is isomorphic to $\Psi$, $p(\lambda \vert P_\psi)=\delta(\lambda-\lambda_\psi)$.''}
\end{quote}

The conception of completeness is deeply related to the conception of reality, since in a classical picture reality is mapped by correspondence rules to a physical theory. Since we expect this theory to describe the whole phenomena we consider it to be complete in the sense of \textit{theory completeness} and by \textit{bijective completeness} we expect it to map the whole of reality. \newline

We conclude that \textit{bijective completeness} is an inherent property of the conception of a classical realism. \textit{Theory completeness} motivates the aim for a theory of everything which is fully complete in the sense of \textit{theory completeness}, and by \textit{bijective completeness} explores and defines all branches of reality. 

\newpage

\section{Reality in contemporary philosophy of science}

A thorough discussion on reality in philosophy of science would go beyond the scope of this work, so we confine our presentation to conceptions that are already related to the debates on the interpretations of Quantum Mechanics. We also omit the historical development, since nearly every philosophical attitude advocated a position towards reality. 

\subsection{Metaphysical realism}
\label{sec_Metaphysical realism}

Metaphysical realism is the position that the world exists independently of us, our knowledge about the world and our perception of it. 

We define a set of elements of reality $ PR $.

\begin{equation}
PR = \{\epsilon_i\} = \{ \epsilon_A,\epsilon_B, \epsilon_C \}
\label{Definition Elements of reality}
\end{equation}

\begin{mydef}
\label{Definition Metaphysical realism}
The world ensues Metaphysical realism if there exist ontological entities $\epsilon_i$ we name elements of reality. 
\end{mydef}

The position of metaphysical realism is controversial and faces a lot of challenges, mainly based on the representation problem, which questions the connection between our beliefs and the independently existing entities in the world. The representation problem is the main attack line for criticism from anti-realists. There are several anti-realist positions that differ vastly, but their common ground is the rejection of the basic assumption of metaphysical realism; the existence of a world independent of our perception and beliefs. \newline

Metaphysical realism is an assumption prior to any statement on the connection to the structure of scientific theories. That is also the basic difference between metaphysical realism and scientific realism; the latter one already involves statements and positions towards the structure of our scientific enterprise. Metaphysical realism does not imply scientific realism \cite[]{sep-realism-sem-challenge}:

\begin{quote}
\textit{``That the world’s constituents exist mind-independently does not entail that its constituents are as science portrays them. One could adopt an instrumentalist attitude toward the theoretical entities posited by science, continuing to believe that whatever entities the world actually does contain exist independently of our conceptions and perceptions of them.''}
\end{quote}

Putnam advances a position of internalist perspective of metaphysical realism and presents the externalist perspective in a brief way \cite[p.48]{putnam1981reason}:

\begin{quote}
\textit{``One of these perspectives is the perspective of metaphysical realism. On this perspective, the world consists of some fixed totality of mind-independent objects. There is exactly one true and complete description of 'the way the world is'. Truth involves some sort of correspondence relation between words or thought-signs and external things and sets of things. I shall call this perspective the externalist perspective, because its favorite point of view is a God's Eye point of view.''}
\end{quote}

\begin{figure}[h]
\centering

\includegraphics[width=0.7\textwidth]{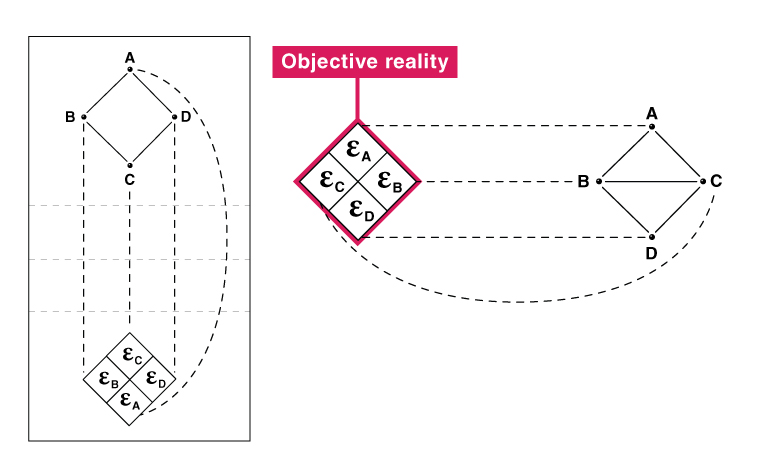}

\caption{Externalist metaphysical realism in the classification scheme}
\end{figure}

It is noteable that Putnam in his presentation of the externalist view amends the basic assumption of metaphysical realism with the assumption of \textit{Bijective completeness}\footnote{See definition \ref{Definition bijective completeness}.} 

Additional to the set of elements of reality $ PR $ we define a set of elements of a theory $ T_{PR}$.

\begin{equation}
T_{PR} = \{ t_i \} = \{ A,B,C \}
\label{Definition Elements of theory}
\end{equation}

\begin{mydef}
\label{Definition Metaphysical realism Externalist view}
The world ensues the externalist perspective of Metaphysical realism if there exist ontological entities $\epsilon_i$ we name elements of reality and there is one mapping $\phi$ that provides a bijection from every element $\epsilon_i$ to $t_i$. 
\end{mydef}

Putnam clearly rejects this externalist view of metaphysical realism and advances a different position \cite[p.49-50]{putnam1981reason}:

\begin{quote}
\textit{``I shall refer to it as the internalist perspective, because it is characteristic of this view to hold that what objects does the world consist off is a question that it only makes sense to ask within a theory or description. Many 'internalist' philosophers, though not all, hold further that there is more than one 'true' theory or description of the world. 'Truth', in an internalist view, is some sort of (idealized) rational acceptability - some sort of ideal coherence of our beliefs with each other and with our experiences as those experiences are themselves represented in our belief system - and not correspondence with mind-independent or discourse-independent 'states of affairs'.''}
\end{quote}

The internalist perspective advances the view that there is a metaphysical reality, but it is only possible only to conclude from the internal perspective of a theory to the nature of the elements of reality. In the internalist perspective, there is not one true and complete description of the world; there are probably more and even equally true descriptions that provide different mappings to the elements of reality. Therefore, we define the internalist perspective as the following:

\begin{mydef}
\label{Definition Metaphysical realism Internalist view}
The world ensues the internalist perspective of Metaphysical realism if there exists ontological entities $\epsilon_i$ we name elements of reality but there is no single mapping $\phi$ that provides a bijection from every element $\epsilon_i$ to $t_i$. There exist several sets $T_{PR-k}$ and corresponding mappings $\phi_k$ from set $T_{PR-k}$ to $PR_k$.
\end{mydef}

The assumption of metaphysical realism amended with statements on the structure of scientific theories represents the conception of scientific realism.

\subsection{Scientific realism}
\label{section_Scientific realism}

Scientific realism is based on metaphysical realism. There are several accounts on scientific realism, and we cannot provide a comprehensive overview of the field. We will follow the position of \cite[]{esfeld2008naturphilosophie}:

\begin{quote}
\textit{`Eine metaphysische Behauptung: Die Existenz und die Beschaffenheit der Welt sind unabh\"angig von den wissenschaftlichen Theorien. \newline
... \newline
Eine semantische Behauptung: Die Beschaffenheit der Welt legt fest welche der wissenschaftlichen Theorien wahr sind (und welche nicht wahr sind).\newline
... \newline
Eine epistemische Behauptung: Die Wissenschaften sind im Prinzip in der Lage, uns einen kognitiven Zugang zur Beschaffenheit der Welt zu gew\"ahren.''}
\end{quote}

\begin{quote}
\textit{``A metaphyiscal proposition: The existence and the nature of the world is independent of scientific theories.\newline
... \newline
A semantic proposition: The nature of the world determines which scientific theories are true (and which are not).\newline
... \newline
An epistemic proposition: Science empowers us to have cognitive access to the nature of the world\footnote{Translation by the author.}.''}
\end{quote}

The first proposition is an addendum to the assumption of metaphysical realism, it is based on the existence of metaphysical realism and demands that scientific theories are independent of this nature of the world. The third proposition assumes that by our best theories we gain access to the metaphysical nature of the world, to the elements of reality. \newline

The conception of scientific realism itself is also a matter of discussion to which we cannot refer in detail in this presentation. An overview on contrasting positions in this question is provided by \cite[]{ellis2005physical}, \cite[]{esfeld2008naturphilosophie}, \cite[]{psillos2005scientific} and \cite[]{sankey2008scientific}. \newline

A contrasting position to scientific realism is the position of Instrumentalism \cite[p.14]{sankey2008scientific}:

\begin{quote}
\textit{``Instrumentalism denies the literal interpretation of theoretical discourse, treating it instead as fictional discourse.''}
\end{quote}

According to instrumentalism it is the aim of science to describe phenomena in a model. Every statement on metaphysical and specifically unobservable entities is superfluous in the framework of science. An instrumentalist's account on Quantum Mechanics is provided by \cite[]{englert2013quantum}:

\begin{quote}
\textit{``Quantum theory has its interpretation and does not need a philosophical ¨ Uberbau. Of course, philosophers should be encouraged to study the lessons of quantum theory — for example, the lesson that the future is not predetermined by the past — but their philosophical debates are really irrelevant for quantum
theory as a physical theory, although they are likely to be of great interest to physicists. We could leave it at that, were there not the widespread habit of the debaters to endow the mathematical symbols of the formalism with more meaning than they have.''}
\end{quote}

The discussion on the definition and relevance of scientific realism raises the question: If unobserved entities exist, what we can know about them? The most prominent discussion on the existence of unobserved entities in physical context is the atomic hypothesis: Do atoms exist? Or to be more precise: Is the conception of an atom within the physical theory necessary? And secondly: Does the conception of atoms imply that the existence of atoms as ontological objects is necessary or at least reasonable? \newline

These two questions refer to different levels, the first only to the conception of a physical theory, the second to the ontological level according to the classification scheme. The question of Atomism should be seen also in the light of the old dichotomy between discrete and continuous. \newline

An historical introduction to Atomism with emphasis on Einstein's contributions is provided by \cite[]{kox2014einstein}, a review on the broader influences and implications of Atomism is provided by \cite[]{roecklein2015locke}, and general introductions on Atomism are provided by \cite[]{chalmers2009scientist} and \cite[]{sep-atomism-modern}. \newline

The acceptance of atoms on the conceptual level was achieved relatively fast, the conception was too successful in the sense of unification and predictive power to be ignored. Therefore, the question arises: What are the principles that lead to the acceptance of new conceptions or ontologies? \newline

This question addresses the distinction between physics and metaphysics which is a heavily disputed topic.\cite[Chapter 4]{sep-metaphysics} gives an overview on the problem and its recent status of the discussion:

\begin{quote}
\textit{``One attractive strategy for answering these questions emphasizes the continuity of metaphysics with science. On this conception, metaphysics is primarily or exclusively concerned with developing generalizations from our best-confirmed scientific theories. For example, in the mid-twentieth century, Quine (1948) proposed that that the “old/intermediate” metaphysical debate over the status of abstract objects should be settled in this way. He observed that if our best scientific theories are recast in the “canonical notation of (first-order) quantification” ..., then many of these theories, if not all of them, will have as a logical consequence the existential generalization on a predicate F such that F is satisfied only by abstract objects. It would seem, therefore, that our best scientific theories “carry ontological commitment” to objects whose existence is denied by nominalism.''}
\end{quote}

There is not one clarified and for all times fixed methodology to gain access to the metaphysical nature of reality; it is more a freedom of methods, which also was supported by \cite[]{feyerabend1993against}. One ontological conception which was both successful and comprehensible was the conception and ontology of objects. 

\newpage

\subsection{Object conception and ontology}

The conception of objects is not a trivial one, but for our discussion the following outline will be sufficient: \newline

An object is an entity that has the ability to carry a set of properties and is distinguished from other objects by this set of properties. It is circumscribed from its environment. \newline

These properties of objects are called intrinsic properties \cite[p.7]{esfeld2007metaphysics}:

\begin{quote}
\textit{``Intrinsic are all and only those properties that an object has irrespective of whether or not there are other contingent objects; in brief, having or lacking an intrinsic property is independent of accompaniment or loneliness.''}
\end{quote}

Properties that are not of that kind are called extrinsic or relational properties. They refer to relations to other objects. \newline

This conception of objects is consistent with Einsteins view on physical objects \cite[]{einstein1948quantum}:

\begin{quote}
\textit{``It is further characteristic of these physical objects that they are thought of as arranged in a space time continuum. An essential aspect of this arrangement of things in physics is that they lay claim, at a certain time, to an existence independent of one another, provided these objects are situated in different parts of space.''}
\end{quote}

Since in classical physics objects like particles, rigid bodies, etc. play an important role in the ontology of these theories, a specific conception of reality is used, which is defined as Entity realism \cite[p.44]{sankey2008scientific}:

\begin{quote}
\textit{``Entity realism is therefore the doctrine that theoretical entities are real in the sense that they exist mind-independently. This may be expressed more simply as
follows: \newline
(ER) The theoretical entities postulated by science are real.''}
\end{quote}

Not all theoretical entities will be regarded as real in the sense of Entity realism. It is a question of the structure of the theory, reason and practicability if a theoretical entity is considered as real. Since objects turn out to be conceptions in the best of our theories (within classical physics), it is reasonable to apply Entity realism to the conception of objects and regard them as real on the ontological level. 
\newpage

There are some arguments against an object ontology that follow directly out of the scientific progress in classical physics. At several states of classical physics and philosophy of nature before, objects on the ontological level have been proposed. During the progress of physics, these object ontologies have been abandoned or modified as long as physics arrived at atoms and elementary particles. All object ontologies that have been adopted on this way, together with respective sets of properties, turned out to have no ontological character. Instead, the properties assigned with those objects followed out further physical conceptions that involved more fundamental ontologies. Applying the Pessimistic Meta Induction to this development process, the object ontology turns out to be not a successful conception, though a helpful one. If object ontology fails on so many levels in classical physics, Pessimistic Meta Induction confined to this question questions why an object ontology should be applied to deeper levels. Would it not be surprising if object ontology failed on all empirical accessible levels and succeeds on the metaphysical level? \newline

Object ontology is deeply related to materialism, since materialism emphasizes the fundamental character of matter as the essential entity of which everything is made. Materialism became challenged by theory of relativity and Atomism which deconstructed the conception of matter as a substance, a process that is continued by elementary particle physics even now. \newline

The emphasis of objects in the conceptual and ontological level of our theories also led to several problems in context with Quantum Mechanics as we will see in Section \ref{sec_ Concepts of reality in QM}. Since the object-conception and ontology is a dominant element in physical theories, we want to present a conception in philosophy of science that challenges the relevance of objects, the conception of structural realism. 

\newpage

\subsection{Structural realism}

Structural realism is a conception in philosophy of science that goes back to \cite[]{worrall1989structural}, who provided a conception for the structure of theories and the involved entities that differs vastly from the approaches in classical physics. The main difference is the type of theoretical entity that structural realism assigns reality to according to Entity realism. Structural realism emphasizes "structure" as the primary ontological entity and reduces the relevance of the entity "object." 

\cite[p.153]{worrall1989structural}:
\begin{quote}
\textit{``Scientific Realism of any stripe consists of a metaphysical thesis conjoined with an epistemological one. The metaphysical thesis is often taken as the claim that there exists a reality independent of the human mind. But, as Elie Zahar points out, a more accurate rendition in our more enlightened, post-Cartesian-dualism times
would be: \newline 
There exists a structured reality of which the mind is a part; and, far from imposing their own order on things, our mental operations are simply governed by the fixed laws which describe the workings of Nature.''}
\end{quote}

and further: 

\begin{quote}
\textit{``Not only is this structured reality partially accessible to human
discovery, it is reasonable to believe that the successful theories
in mature science (the unified theories that explain the
phenomena without ad hoc assumptions) have indeed latched on, in some no doubt partial and approximate way, to that structured reality, that they are, if you like, approximately true.''}
\end{quote}

With the structural realism approach, Worrall attempted to strengthen scientific realism, which had to face main criticism from the Underdetermination argument by \cite[]{laudan1981confutation} and the Pessimistic Meta Induction by \cite[]{quine1951main}. \newline

There are three positions of structural realism that can be distinguished, regarding their way of interpreting the Entity realism:

\begin{itemize}
\item
Epistemic structural realism (ESR) is the position advocated by Worrall emphasizing the structures as primary over objects, but assigning both relevance on the ontological level.
\item
Radical Ontic structural realism (OSR) is the position introduced by \cite[]{ladyman1998-LADWIS}, emphasizing structure, but neglecting objects as superfluous, scrutinizing the metaphysical or epistemological aim of ESR.  
\item
Moderate Ontic structural realism is the position advocated by \cite[]{esfeld2008moderate} putting structures and objects on the same ontological basis. There are no relations without relata.
\end{itemize}

A more detailed overview on the differences of these positions is provided by \cite[]{sep-structural-realism}, an analysis regarding the presented classification scheme and the relation to other approaches of structures of scientific theories is provided by \cite[]{krizek2017classification}. Despite all subtleties ESR and OSR are distinguished by these main positions \cite[]{sep-structural-realism}:

\begin{quote}
\textit{``A crude statement of ESR is the claim that all we know is the structure of the relations between things and not the things themselves, and a corresponding crude statement of OSR is the claim that there are no ‘things’ and that structure is all there is (this is called ‘radical structuralism’ by van Fraassen 2006).''}
\end{quote}

Notwithstanding that object conceptions and ontologies played a crucial role in the conception of classical physics, in Quantum Mechanics the conception of objects becomes challenging and even dubious. We provide involved conceptions of reality in Quantum Mechanics and revisit the relevance of structural realism within Quantum Mechanics.

\newpage

\section{Concepts of reality in Quantum Mechanics}
\label{sec_ Concepts of reality in QM}

The starting point for a new picture of reality is the new formal description of Quantum Mechanics that emerged out of the old Quantum theory and its peculiarities. The formal structure of Quantum Mechanics was different in the sense of the involved framework. It was not defined in a $\mathbb{R}^3 $ space or $\mathbb{R}^{1,3}$ Minkowski spacetime, it was formally defined on a Hilbert space, which has infinite dimensions. The formal structure of states as vectors in a Hilbert space allowed the possibility of superpositions.    

...
ERGÄNZEN die Problematik der Interpretation
...

With respect to that question, there are several conceptions of reality that are provided.

\subsection{Naive realism}

Following the position of metaphysical realism presented in section \ref{sec_Metaphysical realism} there is a reality out there that causes the phenomena we observe. The view of naive realism is that our observations not only resemble the truth about the causes of the phenomena, even more our observations reveal the complete nature of these causes. \newline

In these paradigms observation is a reading off of properties that are out there independently of our observations. Naive realism can involve realism about quantities as numerical representations of the properties of the entities in nature. Further, a naive realism about concepts is conceivable.

\begin{figure}[h]
\centering

\includegraphics[width=0.7\textwidth]{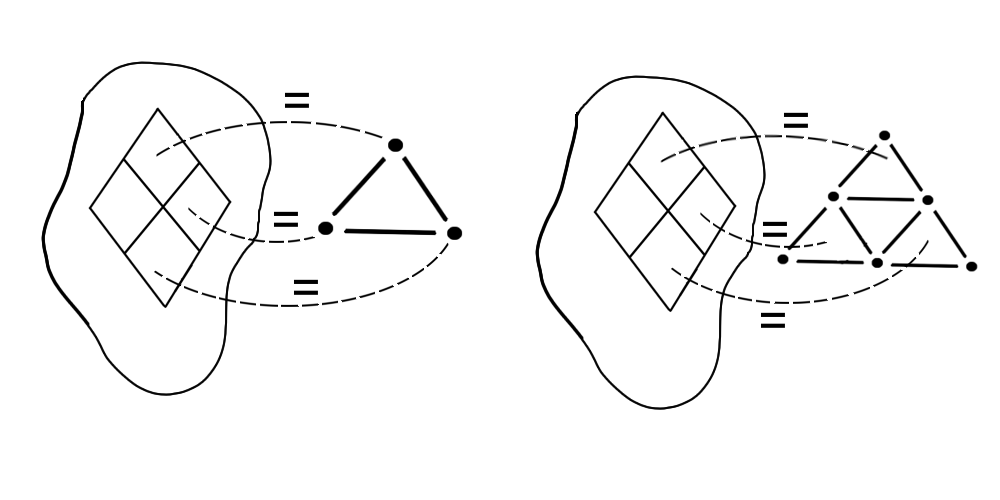}

\caption{Naive realism}
\end{figure}

For a presentation of Einsteins naive realism refer to \cite[]{krizek2017einstein}.

A form of naive realism that is in discussion in contemporary physics is the Mathematical Universe Hypothesis \cite[]{tegmark2004parallel}. It claims that all entities on the ontological level are directly related to the mathematical formalism, and even more are isomorphic to the mathematical formalism. In that sense, the Mathematical Universe Hypothesis trivially explains the unreasonable effectiveness of mathematics and proposed by\cite[]{wigner1960unreasonable}. \newline

Naive realism is a version of the externalist perspective of Metaphysical realism:

\begin{mydef}
\label{Definition Naive realism}
Naive realism is the externalist perspective of metaphysical realism with either conceptual or mathematical elements of the theory being in \textit{bijective completeness} to ontological entities, the elements of reality.
\end{mydef}

Naive realism involves no conception on measurements; therefore, it is not identical to a noninvasive measurability, which will be discussed later. 

\subsection{Einsteins reality criterion}

In the EPR paper \cite[]{einstein1935can}, Einstein presented a criterion for elements of a theory to represent elements of reality. The reality criterion is a methodological claim to identify elements of a theory that fulfil the requirement of being in correspondence with elements of reality. 

Einstein was aware of the metaphysical demarcation problem with physics. In his letter to Schr\"odinger \cite[letter 206, p.537]{vonMeyenn2010entdeckung}, Einstein admits:

\begin{quote}
\textit{``Die eigentliche Schwierigkeit liegt darin, da\ss \, die Physik eine Art Metaphysik ist; Physik beschreibt „Wirklichkeit“. Aber wir wissen nicht, was „Wirklichkeit“ ist; wir kennen sie nur durch die physikalische Beschreibung!''}\footnote{Translation by the author}
\end{quote}

\begin{quote}
\textit{``The difficulty is due to that physics is a kind of metaphysics; physics describes "reality". But we do not know what "reality" is; we only know it through the physical description!''}
\end{quote}

The reality criterion is defined by \cite[]{einstein1935can} in the following way:

\begin{quote}
\textit{``If, without in any way disturbing a system, we can predict with certainty (i.e., with probability equal to unity) the value of a physical quantity, then there  exists an element of physical reality corresponding to this physical quantity.''}
\end{quote}

By definition of a probability measure $P(t_i) $ for the elements of the theory the reality criterion can be formally expressed as: 

\begin{equation}
\forall t_i \in T_{PR} \wedge P(t_i)=1: \exists \epsilon_i \in PR
\end{equation}

For Einstein the reality criterion has no deep ontological meaning or represents a general methodological claim; therefore, he got rid of it in later versions of the EPR argument. \cite[]{fine2009shaky} claims that there are good reasons to believe that the reality criterion was an idea of Podolsky, since it is also incompatible with methodological aspects of Einsteins theory of relativity \cite[]{krizek2017einstein}. 

\newpage

\subsection{Ruarks reality criterion}

In a response to the EPR-paper, Ruark proposed a modification of the reality criterion \cite[p.466]{ruark1935quantum}: 

\begin{quote}
\textit{``This conclusion\footnote{The conclusion that the quantum-mechanical description of the physical reality is complete.} is directly opposed to the view held by many theoreticians, that a physical property of a given system has reality only when it is actually measured, and that wave mechanics gives a faithful and complete description of all that we can learn from measurements. ... The reason\footnote{The reason for the divergence of conclusion.} is that the authors adopt a criterion of reality different from the one just mentioned;''}
\end{quote}

And further: 
\begin{quote}
\textit{``The conclusion drawn by Einstein, Podolsky and Rosen is, that in accordance with their criterion of reality, the quantities corresponding to both P and Q possess reality; so that the wave description, failing to yield both of them, is incomplete.''}
\end{quote}

According to Ruark, none of both quantities correspond to an element of reality unless measured. This modified reality criterion resembles the position of Bohr, but without any commitment to classicality of the measurement apparatus. 

\subsection{Local realism in the Bell inequalities}

Bell proposed an inequality in \cite[]{bell1964a} that demonstrates the incompatibility of Quantum Mechanics with a set of assumptions in classical physics. These assumptions are locality and realism. We want to give a closer look to the definition of the latter one. \newline

In its original intention Bell's theorem should demonstrate that only non-local hidden-variable theories like the De-Broglie-Bohm interpretation are not in contradiction with Quantum Mechanics, and all local hidden variable theories are ruled out, serving as an argument for the De-Broglie-Bohm interpretation \cite[]{bell1964a}:

\begin{quote}
\textit{``... a hidden variable interpretation of elementary quantum
theory has been explicitly constructed. That particular interpretation has indeed a grossly nonlocal structure. This is characteristic, according to the result to be proved here, of any such theory which reproduces exactly the quantum mechanical predictions.''}
\end{quote}

It is irony that Bell's theorem later was used the other way around to argue that hidden-variable theories are untenable, and only by admitting that the De-Broglie-Bohm interpretation as a non-local hidden-variable theory, it was not affected by Bell's theorem. \newline

Bell proposes the Bohm-version of the EPR argument with a pair of Spin-$\frac{1}{2}$ particles in a singlet state \cite[]{bell1964a}: 

\begin{quote}
\textit{``Since we can predict in advance the result of measuring any chosen component
of $\vec \sigma_2$' by previously measuring the same component of $\vec \sigma_1$ , it follows that the result of any such measurement must actually be predetermined. Since the initial quantum mechanical wave function does not determine the result of an individual measurement, this predetermination implies the possibility of a more complete specification of the state. Let this more complete specification be effected by means of parameters $\lambda$.''}
\end{quote}

The parameter $\lambda$ is the hidden variable that affects the value of the Spin vectors. In this classical model spin is assumed as a classical object property of the particles \cite[p.1]{bell1981bertlmann}:

\begin{quote}
\textit{``Do we not simply infer that the particles have properties of some kind, detected somehow by the magnets, chosen ala Bertlmann\footnote{Bell here refers to the Bertlmann's socks correlation as an example for a classical correlation.} by-the Source - differently for the two particles?''}
\end{quote}

Considering this, the conception of realism in Bell's theorem can be defined as:

\begin{mydef}
\label{Definition Bell Realism}
Bell realism is the conception that a Bell-realistic theory describes objects with properties that are well-defined before the measurement process and after it.  
\end{mydef}

Since the EPR argument was in the first place a way to criticise the statistical character of Quantum Mechanics by isolating the systems on hand of the principle of separability, Quantum Mechanics and, specifically the feature of entanglement, turned out to disobey the principle of separability. Soon the discussion on the EPR argument turned and separability was not longer an argument to isolate the measurement processes of the two constituents involved in a singlet state. Even more separability and locality itself were challenged. The only escape out of this situation from a classical realism perspective is to attribute correlation properties to the entangled constituents of the system \cite[p.3]{bell1981bertlmann}:

\begin{quote}
\textit{``To avoid such action at a distance they have to attribute, to the space-time regions in question, real properties in advance of observation, correlated properties, which predetermine the outcomes of these particular observations. Since these real properties, fixed in advance of observation, are not contained in quantum formalism, that formalism for EPR is incomplete.''}
\end{quote}

What Bell's theorem shows is that such an assignment of local pre-existing properties to objects does not reproduce the results as obtained from Quantum Mechanics. And the possibility to perform experiments to test if nature behaves according to these local pre-existing properties or according to Quantum Mechanics excels the value of Bell's theorem. \newline

A huge number of experiments have been performed to test Bell's theorem and they convincingly demonstrated that nature behaves according to Quantum Mechanics \cite[]{clauser1969Bellexperiment},\cite[]{weihs1998violation}. There are a number of loopholes in the conception of the experiments or the conception of Bell's theorem that could save local Bell-realistic theories. In the last decades, several of them were closed, but a full loophoole-free experiment has not been done.  \newline

Concluding, we can say that Bell-realism is defined by the object character of the quantum system carrying properties. In the classification scheme, this can be illustrated in the following way. \newline 

In a Bell-realistic theory, there are elements of the theory on level one and level two of the classification scheme. Corresponding to the elements of the theory, there are objects as entities on the ontological level that carry properties. 

\begin{figure}[h]
\centering

\includegraphics[width=0.9\textwidth]{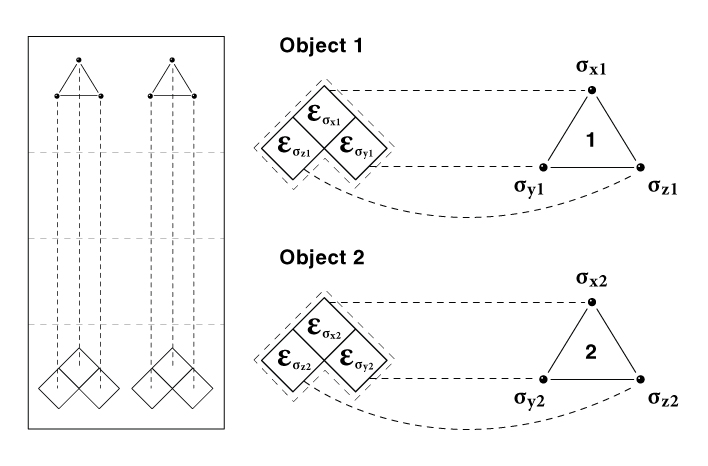}

\caption{Bell realism in the Classification scheme}
\end{figure}
 
 \newpage 
 
\subsection{Reality in the measurement process}

In Quantum Mechanics, the measurement postulate is needed to explain how single eigenvalues arise out of the formalism, which describes a state generally in a superposition of several eigenfunctions and their according eigenvalues \cite[]{von1955mathematical}.

For classical physics, we want to propose a reality conception that describes the measurement process in classical terms.

\begin{mydef}
\label{Definition Naive Measurement realism}
Naive Measurement Reality is the conception that on the ontological level
\begin{compactenum}[a)]
\item objects exist,
\item that carry a set of properties,
\item whose values are independent of the measurement process.
\end{compactenum}
\end{mydef}

The difference between Bell realism and Naive Measurement realism is the ontological commitment to the existence of objects of the latter one. Bell realism does not involve such an assumption. \newline

In the conception of Naive Measurement Reality, a measurement of the property of an object means that the value of the property of the object remains unaltered by the measurement process. The system that conducts the measurement changes one of its properties by the measurement process in a way that its value corresponds to or equals the value of the property of the measured system. \newline

On the ontological level, the measured systems property is an element of the theory and corresponds to an element of reality; together they fulfil \textit{bijective completeness}. By the measurement process, as explained above, one value of the measuring system will become corresponding or equal to the property of the measured system. By this measurement process, the property of the system that conducts the measurement alters one of its elements of reality. After the measurement, the measuring system will also fulfill \textit{bijective completeness}. \newline

The Naive Measurement realism proposes a noninvasive measurability, as the system that has been measured remains in its state unaltered, and a corresponding or equal value on the measurement apparatus evolves out of the measurement process. \newline 

If we apply this conception to the measurement process and the mathematical formalism in Quantum Mechanics, several problems arise. \cite[]{gardner1972quantum} provides an analysis on the three positions of Bohr, Popper and Reichenbach and their relation to the Kochen-Specker theorem. We provide an overview of the reality conceptions in the picture of the classification scheme on these three positions. \newpage

\subsubsection{The Bohr view}

Despite all subtleties of Bohr's view on Quantum Mechanics we want to discuss his ideas on the conception of reality at hand of the measurement process. An thorough overview on Bohr's positions is provided by \cite[]{kragh2012niels}.\newline

In Bohr's view, the measurement process is seen as a creation process, where by an interaction of a microscopic system with a macroscopic system, an element of reality for the microscopic system in the macroscopic regime is created. \newline

In the formalism of Quantum Mechanics, the state of the microscopic system can be written by a decomposition to eigenvectors and eigenstates regarding the observable $A$: 

\begin{equation}
\ket{\psi}=\sum_{i}a_i\ket{\phi_i}
\end{equation}

\begin{equation}
A\ket{\phi_i}=a_i\ket{\phi_i}
\end{equation}

The expectation value of $A$ is given by

\begin{equation}
\expval{A}=\expval{A}{\Psi}=\sum_{i}a_i|\bra{\psi}\ket{\phi_i}|^2.
\end{equation}

Bohr's view on Quantum Mechanics involves a discrimination between the microscopic world and the macroscopic world on the ontological level, though he did not provide a criterion that gave a sharp distinction. In this view, Quantum Mechanics describes the outcomes of interactions with the microscopic world. 

A measurement process involves a classical apparatus with properties $\{t_i\}=\{M_1,M_2,...,M_i\}$ which has corresponding elements of reality $\{PR\}=\{\epsilon_1,\epsilon_2,...,\epsilon_i\}$. Measuring the observable $A$ will reduce the state vector according to the measurement postulate to one of its eigenvectors. 

\begin{equation}
\ket{\psi}=\sum_{i}a_i\ket{\phi_i} \Longrightarrow \ket{\psi}=a_j\ket{\phi_j}
\end{equation}

By this interaction with the classical measurement apparatus, one of the properties of the measurement apparatus will alter its value to the value of the respective eigenvalue. \newline

The microscopical system will be in the reduced state and a repetitive measurement will repeat the measurement result. Therefore, it is seen as justified to assign to the element of the theory $a_j$ an element of reality, and the elements of reality obey \textit{bijective completeness}. \newline

In Bohr's view, it is essential that we cannot speak about the nature of the microscopic entities before there has been a measurement; unmeasured observables have no meaning. In that way Bohr argued in his reply against Einsteins presuppositions in the EPR paper\cite[]{bohr1935can}.

\begin{mydef}
\label{Definition Bohr realism}
Bohr Reality is the conception that on the ontological level it is meaningless to speak about entities of the microscopic regime. It is only meaningful to speak about macroscopic entities.
\end{mydef}

\begin{figure}[h]
\centering

\includegraphics[width=0.8\textwidth]{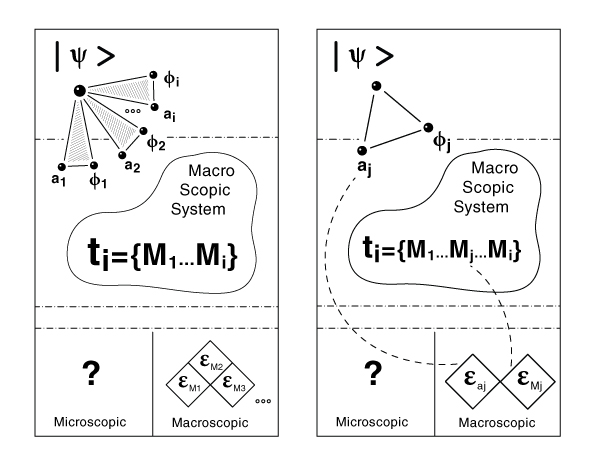}

\caption{Bohrs view of a measurement - before and after the measurement}
\end{figure}

\subsubsection{Popper's view}

Popper's view is distinguished from Bohr's view by a few differences. The notion of a microscopic and macroscopic world is not a core conception as in Bohr's view. To Popper, a quantum mechanical system always has to have properties even before the measurement. In Popper's view, there exists an element of reality that cannot be assigned to one of the elements in the theory. It is more or less attributed to the conception of the particle, but not accessible within the theory. \newline

After the measurement, there is an element in the theory that corresponds to an element of reality. In general, it will not be the same element of reality, since the property might have changed.

What is essential to Popper, according to Gardener, is the assignment to values before the measurement process \cite[]{gardner1972quantum}:

\begin{quote}
\textit{``He simply wishes to claim that it is at least possible (i.e., consistent with quantum theory) to maintain that unmeasured observables have values. A realistic view need not to be subject to the same strictures as an avowedly deterministic one.''}
\end{quote}

And in more detail concerning the elements of the theory and the elements of reality \cite[]{gardner1972quantum}: 

\begin{quote}
\textit{``Another way to state Popper's view, taking cognisance of Kochen and Specker, might run as follows. We must distinguish between (1) quantitative properties (or physical magnitudes), such as spin, position, etc., and (2) the corresponding observables which go under the same names. A value of an observable is by definition the result of a measurement of that observable or, equivalently, of the corresponding quantitative property. When a quantitative property is measured, its value always coincides with that of the corresponding observable. However, at other times, the quantitative properties may have values which they never have when measured because the algebraic structure of the observables prevents the corresponding observables from having those values.''}
\end{quote}

What is most disturbing in Popper's view is that the assignment of the ontological elements of reality changes during the measurement process in a way that affects the nature of the correspondence between elements of the theory level and ontological elements. This is not satisfying from the perspective of the classification scheme.

Since Popper's view failed to be accepted and in some sense could be seen as a special case of the De Broglie-Bohm account, we will not provide an extra definition for this conception of reality.

\begin{figure}[h]
\centering

\includegraphics[width=0.8\textwidth]{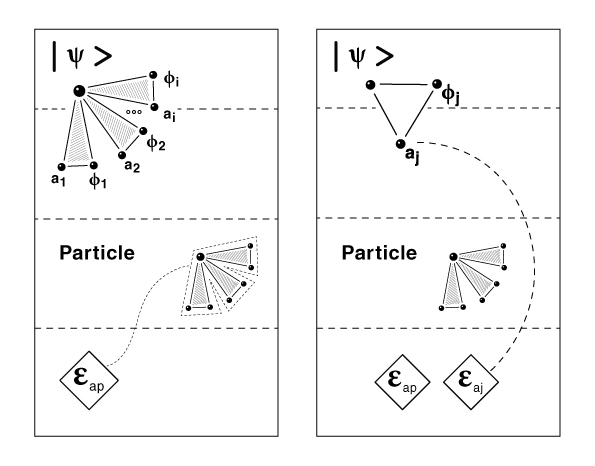}

\caption{Popper's view of a measurement - before and after the measurement}
\end{figure}

\subsubsection{Reichenbach's view}

The conception of Reichenbach is based on critique of the Copenhagen interpretation of Quantum Mechanic. Reichenbach also refers to the completeness argument by \cite[]{einstein1935can}, and a rejection of Bohr's approach of meaninglessness of entities in the microscopical world \cite[p.22ff]{reichenbach1944philosophic}:

\begin{quote}
\textit{``Although we should like to consider this Bohr-Heisenberg interpretation as
ultimately correct, it seems to us that this interpretation has not been stated
in a form which makes sufficiently clear its grounds and its implications. In the
form so far presented it leaves a feeling of uneasiness to everyone who wants
to consider physical theories as complete descriptions of nature; the path
towards this aim seems either to be barred by rigorous rules forbidding us to
ask questions of a certain kind, or open only to vague pictures which have no
claim to be regarded as an adequate expression of reality.''}
\end{quote}

As a solution he offers a three-valued logic that in some sense obeys Bohr's ideas of undefined observables before the measurement. But in Reichenbach's view, this does not mean that it is impossible to speak about the ontological entities of the theory \cite[p.43]{reichenbach1944philosophic}:
\begin{quote}
\textit{``We have, therefore, good reasons to say that the language of quantum mechanics is written in terms of a three-valued logic. We must not forget, however,
that the subject matter of a science, without the addition of further
qualifications, does not determine a particular form of logic. Quantum mechanics
can be constructed in the form of a two-valued logic; this is demonstrated
by the existence of exhaustive interpretations. Only when we introduce
the postulate that causal anomalies be not derivable are we obliged to turn to
a three-valued logic.''}
\end{quote}

The three-valued logic allows true, false, and indeterminate statements \cite[p.148]{reichenbach1944philosophic}:

\begin{quote}
\textit{``This conception is introduced by a definition concerning unobserved entities
differing from definition 1\footnote{Which defines basically a conception that resembles Naive measurement realism.} with respect to the rule which establishes the values existing before the measurement. This is the following definition.''}
\end{quote}

\begin{quote}
\textit{``Definition 3. The value of an entity $u$ measured in a situation s means the value $u$ existing after the measurement; before the measurement, and thus in the situation s, the entity $u$ has all its possible values simultaneously.''}
\end{quote}

\begin{quote}
\textit{``By possible values we mean values which in the situation s can be expected with a certain probability greater than to be the result of a measurement.''}
\end{quote}

By "indeterminate," Reichenbach means that in the picture of the classification scheme that there are all possible entities in existence. Since Reichenbach speaks explicitly about entities carrying values, he advanced a kind of mathematical realism. \newline

After the measurement, one of the entities values is selected by the measurement postulate and the theory element $a_j$ is in correspondence to the element of reality $\epsilon_{aj}$ and fulfils \textit{bijective completeness}.

\begin{figure}[h]
\centering

\includegraphics[width=0.8\textwidth]{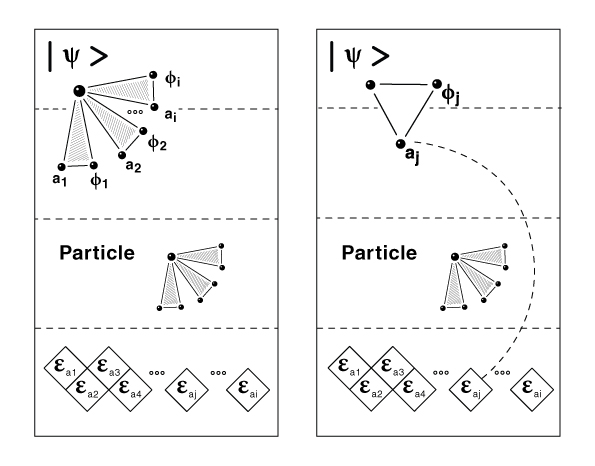}

\caption{Reichenbach's view of a measurement - before and after the measurement}
\end{figure}

Reichenbach's ideas on a modification of logical presuppositions to give an appropriate interpretation found its continuation in the Quantum Logic approaches \cite[]{mittelstaedt1978quantum}, the idea of many possible values that exist simultaneously resembles the Many-World approaches in the interpretation of Quantum Mechanics. 

\newpage

\subsection{Structural realism in Quantum Mechanics}

Already, \cite[p.123]{worrall1989structural} recognized the implications of the structural realism approach for Quantum Mechanics:

\begin{quote}
\textit{``Is there any reason why a similar structural realist attitude cannot be
adopted towards quantum mechanics? This view would be explicitly divorced
from the ‘classical’ metaphysical prejudices of Einstein: that dynamical
variables must always have sharp values and that all physical events are fully
determined by antecedent conditions. Instead the view would simply be that
quantum mechanics does seem to have latched onto the real structure of the
universe, that all sorts of phenomena exhibited by microsystems really do
depend on the system’s quantum state, which really does evolve and change in the way quantum mechanics describes.''}
\end{quote}

Or in a more compact way \cite[p.123]{worrall1989structural}:

\begin{quote}
\textit{``The structural realist simply asserts, in other words, that, in view of the
theory’s enormous empirical success, the structure of the universe is (probably) something like quantum mechanical.''}
\end{quote}

Referring to the ontological level, structural realism rejects the idea of Bohr that statements and questions about the microscopical world are meaningless, but structural realism advances a realist view far from naive realism, related to scientific realism \cite[p.122]{worrall1989structural}:

\begin{quote}
\textit{``The realist is forced to claim that quantum-mechanical states cannot be
taken as primitive, but must somehow be understood or reduced to or defined in classical terms.
But the structural realist at least is committed to no such claim - indeed
he explicitly disowns it. He insists that it is a mistake to think that we can ever
“understand” the nature of the basic furniture of the universe.''}
\end{quote}

What is emphasized by structural realism is the primary role of relations instead of a primary role of objects, but to which extent is a matter of dispute within the structural realist movement. \cite[]{Ladyman1998409} first suggested that the position of \cite[p.123]{worrall1989structural} is ambiguous and introduced a distinction in epistemic and ontic structural realism, advancing the latter position. Worrals  epistemic structural realism position is settled in the conception of Conservatism  \cite[p.101]{morganti2004preferability}:

\begin{quote}
\textit{``... the proper interpretation of ESR is, I believe, exactly that according to which we are rationally entitled to believe in some level of preservation of structures (more or less faithfully representing reality) in the history of science but, considering the uncertainty existing at the level of our tentative ontological description of the universe, we should be 'agnostic' about what exists beyond those structures''}
\end{quote}

In Worrals view the objects as entities connected by structures are part of the structures, but not the primary ones. Ladymans positions is nowadays named as radical ontic structural realism and characterised by its full rejection of the concept of object entities, or individual entities \cite[p.422]{Ladyman1998409}:

\begin{quote}
\textit{``The demand for an individuals-based ontology may be criticised on the grounds
that it is the demand that the structure of the mind-independent world be imaginable
in terms of the categories of the world of experience. Traditional realism should
be replaced by an account that allows for a global relation between models and
the world, which can support the predictive success of theories, but which does
not supervene on the successful reference of theoretical terms to individual entities,
or the truth of sentences involving them.''}
\end{quote}

Related to this radical ontic structural realism, there is a position of moderate ontic structural realism advanced by \cite[]{esfeld2008moderate} which is situated somewhere in between ontic and epistemic structural realism. The difference is again the ontological relevance of objects; to moderate structural realism, objects are an on the same footing as structures, in the sense that there are no relations without entities thate those relations refer to. In the context of Quantum Mechanics, moderate structural realism takes the following position \cite[p.8]{esfeld2007metaphysics}:

\begin{quote}
\textit{``Quantum physics can therefore be taken to suggest a metaphysics of relations, known also as structural realism: insofar as quantum physics is concerned, the fundamental physical properties consist in certain relations instead of being intrinsic properties.''}
\end{quote}

Moderate structural realism on one hand emphasizes the primary role of relations, but without abandoning the conception of objects in general \cite[p.11]{esfeld2004quantum}:

\begin{quote}
\textit{``The argument of this paper accepts that relations require things that stand in the relations (although these things do not have to be individuals, and they need not have intrinsic properties) and regards physical theories as referring to things.''}
\end{quote}

The merit of the radical OSR position following \cite[p.29]{esfeld2013ontic} is ...

\begin{quote}
\textit{``... the commitment to there being a certain sort of structures in the domain of quantum physics instead of objects with an intrinsic identity.''}
\end{quote}

What Esfeld means exactly is that out of the formalism of Quantum Mechanics and its interpretation, and the framework of ontic structural realism, it is conclusive that there are no objects with intrinsic properties on the ontological level. What ontic structural realism is abandoning is the conception of objects as entities on the ontological level of the classification scheme, replacing it with the notion of structures. \newline

What goes beyond this insight, is the claim that OSR cannot solve the problem without involving the interpretational positions of Quantum Mechanics \cite[p.29]{esfeld2013ontic}:

\begin{quote}
\textit{``OSR is not in the position to provide on its own an ontology for QM, since it does not reply to the question of what implements the structures that it poses.''}
\end{quote}

With this conclusion \cite[]{esfeld2013ontic} presents an overview on three main candidates for interpretations of Quantum Mechanics that provide ontological implications: Many worlds interpretations, collapse interpretations, and hidden variable interpretations. Each of these interpretations provides a different candidate for the structure in the ontic structural realism program, and thereby completing OSR with an concrete structure ontology. \newline

\cite[p.26]{esfeld2013ontic} refers to Maudlin and presents three propositions for the measurment problem:

\begin{quote}
\textit{``1.A The wave-function of a system is complete, i.e. the wave-function specifies(directly or indirectly) all of the physical properties of a system. \newline
1.B The wave-function always evolves in accord with a linear dynamical equation
(e.g. the Schrödinger equation). \newline
1.C Measurements of, e.g., the spin of an electron always (or at least usually) have
determinate outcomes, i.e., at the end of the measurement the measuring device
is either in a state which indicates spin up (and not down) or spin down (and not
up).''}
\end{quote}

The three main interpretations according to \cite[]{esfeld2013ontic} all give up one of these three propositions. Hidden variable interpretations give up 1A, collapse interpretations give up 1B and many worlds-interpretations give up proposition 1C. \newline

What is omitted in \cite[]{esfeld2013ontic} are several other approaches in the interpretations of Quantum Mechanics. We want to amend the presentation of \cite[]{esfeld2013ontic} by three more interpretational approaches that emphasize the conceptions of information and belief, but in completely different directions.

\newpage

\subsubsection{De Broglie-Bohm interpretation}

The De Broglie-Bohm interpretation is distinguished in essential aspects from Bohmian mechanics in the reading of \cite[]{durr2009bohmian}. It is characterized by the view that the Schr\"odinger equation can be rewritten as a set of a complex- and a real-valued equation and resembles a Hamilton-Jacobi equation known from classical mechanics, supplemented by a term that has no correspondence in classical physics, the Quantum potential. Proponents of the De Broglie-Bohm interpretation emphasize neither the quasi-Newton character of the equations nor a return to classical physics. Even more, they emphasize the highly nonclassical character of the Quantum potential as an expression of a completely different physics. Therefore, it shows a deep misunderstanding of the De Broglie-Bohm interpretation if we follow the argumentation in \cite[p.28]{esfeld2013ontic}:

\begin{quote}
\textit{``In enquiring whether OSR can be applied to Bohm’s theory, let us focus on the contemporary conceptualization of this theory known as Bohmian mechanics, which
abandons the quasi-Newtonian formulation of the theory in terms of a quantum potential''}
\end{quote}

For Bohmian Mechanics, \cite[]{esfeld2013ontic} identifies a candidate for the structure of a structural realism as: 

\begin{quote}
\textit{``it is a structure that takes all the particles as relata. In short, on Bohmian mechanics, there is a structure of correlated particle positions.''}
\end{quote}

In the De Broglie-Bohm interpretation this structure that takes all particle positions as relata in a very natural way, directly out of the reinterpretation of the Schr\"odinger equation, is the Quantum potential. For k particles it is defined as

\begin{equation}
Q = \sum_{k=1}^{N} - \frac{\hbar^2\nabla^2_k R}{2m_kR}.
\label{equ_Quantum potential}
\end{equation}

The Quantum potential depends on a differential operator applied to the real-valued function $R$ with the polar decomposition $ \psi=R e^{i\frac{S}{\hbar}} $. The real-valued function $R$ depends in general on the position of all k particles:

\begin{equation}
R = R(\vec{q}_1,...,\vec{q}_k,t)
\label{equ_real valued function}
\end{equation}

Thereby, the De Broglie-Bohm interpretation also instantiates the structure for a structural realist approach, but in the case of the hidden variable approach, in a more transparent way, with an explicit mathematical object, than in Bohmian mechanics. \newpage

Proponents of the De Broglie-Bohm interpretation emphasize the Quantum potential as an information potential, but with a different conception of information than in other informational approaches in Quantum Mechanics \cite[p.90]{holland1995quantum}; \cite[p.260]{bohm1984measurement}\footnote{

It should be noted further that Bohm emphasized the preliminary character the Quantum potential has to him\cite[p.102]{bohm1981wholeness}: 

\begin{quote}
\textit{``Nevertheless, we evidently cannot be satisfied with accepting such a potential in a definitive theory. Rather, we should regard it as at best a schematic representation of some plausible physical idea to which we hope to advance later, as we develop the theory further.''}
\end{quote}
}. \newline

It is information in the sense of in-formation\footnote{This reading of in-formation refers to Aristotle's formal cause.}, not in the sense of knowledge of a system perceiving its environment\cite[9.327]{bohm1987ontological}: 

\begin{quote}
\textit{``In this explanation of the quantum properties of the electron, the notion of information plays a key role. Indeed it is helpful to extend this notation of information and introduce what could be called active information. The basic idea of active information is that a form having very little energy enters into and directs a much greater energy. The activity of the latter is in this way given a form similar to that of the smaller energy. It is therefore clear that the original energy-form will “inform” (i.e. put form into) the activity of the larger energy.''}
\end{quote}

In the picture of Maudlins propositions on the measurement problem, the De Broglie-Bohm interpretation gives up the proposition 1A, in the sense that it interprets the real-valued equation part of the Schr\"odinger equation as a Hamilton Jacobi function and, as a consequence of this, an additional equation of motion for the particle positions follows. \newline

Proponents of the De Broglie-Bohm interpretation would identify active information and its mathematical carrier, the Quantum potential, as the resource that represents the structure that is requested by a structural realist approach. 

\newpage

\subsubsection{Brukner-Zeilinger interpretation}

The Brukner-Zeilinger interpretation has been proposed as a foundational principle to reconstruct Quantum Mechanics out of informational principles \cite[]{zeilinger1999foundational}. \newline

For a presentation of the Brukner-Zeilinger interpretation refer to \cite[]{brukner1999operationally}; \cite[]{brukner2003information}; \cite[]{kofler2013quantum} and \cite[]{zeilinger1999foundational}. For critical remarks refer to \cite{timpson2003applicability}; \cite{timpson2013quantum}; \cite[]{khrennikov2016reflections} and \cite[]{khrennikov2012vaxjo}. \newline

Following the ideas of the Brukner-Zeilinger interpretation, information is seen as the fundamental entity on the ontological level. \newline

\cite[p.642]{zeilinger1999foundational}:
\begin{quote}
\textit{``The most fundamental viewpoint here is that the quantum is a consequence of what can be said about the world. Since what can be said has to be expressed in propositions and since the most elementary statement is a single proposition, quantization follows if the most elementary system represents just a single proposition.''}
\end{quote}

By decomposition of complex systems into smaller and smaller parts one ends up at systems where such a single proposition represents the whole information content of the system. These type of systems are elementary systems according to \cite[p.635]{zeilinger1999foundational}. \newline

\cite[p.635]{zeilinger1999foundational} emphasizes that this foundational postulate can be seen in a rather instrumentalist view \cite[p.635]{zeilinger1999foundational}:

\begin{quote}
\textit{``I would like to underline that notions such as that
a system ``represents’’ the truth value of a proposition or that it ``carries’’ one bit of information only implies a statement concerning what can be said about possible measurement results.''}
\end{quote}

To delimit this informational approach from subjective approaches like Qbism \cite[]{fuchs2016qbist} or the Växjo interpretation \cite[]{khrennikov2012vaxjo}, Zeilinger emphasizes that information is not completely replacing the notion of reality, but he enhances the ontological status of information \cite[p.642]{zeilinger1999foundational}:

\begin{quote}
\textit{``It is evident that one of the immediate consequences is that in physics we cannot talk about reality independent of what can be said about reality. Likewise it does not make sense to reduce the task of physics to just making subjective statements, because any statements about the physical world must ultimately be subject to experiment.
Therefore, while in a classical world-view, reality is a primary concept prior to and independent of observation with all its properties, in the emerging view of quantum mechanics the notions of reality and of information are on an equal footing.''}
\end{quote}

Contrary to the previously mentioned subjective approaches, the Brukner-Zeilinger interpretation makes use of the conception of a hypothetical observer to omit the assignment of the quantum state to one specific observer. This inter-subjective character of information and the description by Quantum Mechanics is emphasized by \cite[p.6]{brukner2017quantum}:

\begin{quote}
\textit{``The quantum state is a representation of knowledge necessary for a hypothetical observer – respecting her experimental capabilities – to compute probabilities of outcomes of all possible future experiments.''}
\end{quote}

In context of David Deutsch's version of the Wigner's friend Gedankenexperiment, Brukner argues for the existence of facts only in a relative sense \cite[p.1]{brukner2017quantum}:

\begin{quote}
\textit{``The view that holds the coexistence of the “facts of the world” common both
for Wigner and his friend runs into the problem of the hidden variable program.
The solution lies in understanding that “facts” can only exist relative
to the observer.''}
\end{quote}

In that sense, reality and facts about the world are defined in a relational way only. This is supported by structural realism, which would emphasize this relational structure. Unclear remains the status of the quantum objects. The Brukner-Zeilinger interpretation remains neutral on the ontological structure of the objects on the quantum level, in a general sense, but it abandons the idea of classical pre-existing properties above the informational content of the elementary systems. \newline

It is this informational storage capacity of an elementary system and its contained total information that is the candidate for a structure according to structural realist ideas. Entanglement, then, is the practical feature of quantum systems that share a structure of common or joint information.  \newline

One aspect of the Brukner-Zeilinger interpretation is the relation between the definition of information and the ways humans process information, as pointed out by  \cite[p.8]{khrennikov2016reflections}:

\begin{quote}
\textit{``I think that to explain peculiarity of the quantum representation of information one has to discuss the logical structure of information processing by humans. Besides classical Boolean logic, there exist various nonclassical logics, nonclassical rules for information processing.
...
Meanwhile, the brain may use more complex logical systems. In particular, it may use quantum logic. Thus the quantum representation of information is a mathematical signature of the use of quantum logic in reasoning.''}
\end{quote}

Despite the inter-subjective approach in the Brukner-Zeilinger interpretation, Khrennikov emphasizes the subjective character of the whole process of information processing and the various definitions of logic. Contrary to the De Broglie-Bohm interpretation, where information is understood as a formal cause, information in the Brukner-Zeilinger interpretation is a still an entity that is related to observers, to humans. \newline

\subsubsection{QBism}

A third approach which we want to relate to structural realism is QBism. QBism, earlier an abbreviation for Quantum Bayesianism, now renamed to Quantum Bettabilitarianism \cite[]{fuchs2016qbist}, is a continuation of the Copenhagen interpretation. It is a reconstruction of Quantum Mechanics that emphasizes the subjective character of Quantum Mechanics, or merely the mixture of subjective and objective elements in one formalism. According to the proponents of QBism, this mixture is the source of the difficulties with the interpretation of the formalism of Quantum Mechanics \cite[p.8]{fuchs2017notwithstanding}:

\begin{quote}
\textit{``The starting point was, along with Jaynes, that the objective could never reside in the probabilities themselves; that was for the subjective. The objective must be somewhere else in the theory.''}
\end{quote}

A reconstruction of Quantum Mechanics based on SIC-POVMs, which are symmetric, informationally complete, positive operator-valued measures, is proposed to decompose the mixture of subjective and objective elements into a different formalism that emphasizes the probabilistic and subjective character of an agent interacting with nature. QBism speaks also about objective elements, which are identified as the Hilbert space dimension of the quantum systems \cite[p.47]{fuchs2016qbist}:

\begin{quote}
\textit{``Quantum theory|that user's manual for decision-making agents immersed in a world of some yet to be fully identified character makes a statement about the world to the extent that it identifies a quality common to all the world's pieces. QBism says the quantum state is not one of those qualities. But of Hilbert spaces themselves, particularly their distinguishing characteristic one from the other, dimension,... QBism carries no such grudge. Dimension is something one posits for a body or a piece of the world, much like one posits a mass for it in the Newtonian theory. Dimension is something a body holds all by itself, regardless of what an agent thinks of it.''}
\end{quote}

QBism and the Brukner-Zeilinger interpretation agree in the emphasis of the significance of the Hilbert space dimension, but disagree on the interpretation of the very nature of this characteristic. Brukner and Zeilinger assign the conception of information to the elementary system, whereas QBism speaks more generally of a capacity \cite[p.48]{fuchs2016qbist}:

\begin{quote}
\textit{``The claim here is that quantum mechanics, when it came into existence, implicitly recognized a previously unnoticed capacity inherent in all matter call it quantum dimension.''}
\end{quote}

By the emphasis of the Hilbert space dimension, QBism follows an old methodology of philosophy and physics equally; the methodology of identifying invariants as foundational and ontological entities. This invariant is the Hilbert space dimension or quantum dimension of a quantum system. \newline

Following the classification of \cite[]{cabello2015interpretations} on interpretations of Quantum Mechanics, QBism is classified as a Type II interpretation on belief. If we want to relate QBism to structural realism, neither ontic structural realism, nor epistemic structural realism seems suitable to represent the position of QBism, since both structural realist positions do not consider the subjective perspective of an agent in the world as emphasized by QBism. \newline

Therefore we propose QBism as a different form of structural realism, a doxastic structural realism. Doxastic is derived from the greek word doxa, which stands for "belief", and reflects the distinction from epistemic and ontological commitments. The structures of this doxastic structural realism, that is represented by QBism, are the mutually beliefs of agents performing Quantum Mechanics. The structures identified by QBism are not objective, since they represent an internalist view of the beliefs on the behaviour of nature of one specific agent in interaction with nature. \newline

The Qbist position also alters the definition of physics its purpose \cite[p.47]{fuchs2016qbist}:

\begin{quote}
\textit{``What is being realized through QBism's peculiar way of looking at things is that physics actually can be done without any accompanying vicious abstractionism. You do physics as you have always done it, but you throw away the idea everything is made of [Essence X]" before even starting.
Physics in the right mindset is not about identifying the bricks with which nature
is made, but about identifying what is common to the largest range of phenomena it can
get its hands on.''}
\end{quote}

So besides the subjective beliefs of agents, QBism identifies the quantum dimension as some objective structure that can be identified as common to a large range of phenomena. In our view this property reflects some aspect of the nature of the universe. \newline

The identification of the quantum dimension as a capacity or resource entails consequences. One has to deal with infinite dimensional Hilbert spaces.  \cite[p.52]{fuchs2016qbist} speculate that the quantum dimension should be finite in general:

\begin{quote}
\textit{``To make quantum dimension meaningful in ontic terms, as a quality common to all physical objects, is to say it should be finite - going up, going down from this object to the next, but always finite. Every region of space where electromagnetism can propagate, finite. Every region of space where there is a gravitational "field," finite.''}
\end{quote}

Space takes a exceptional position under all degrees of freedom in Quantum Mechanics and presumably the solution to this conundrum lies beyond Quantum Mechanics. By means of different interpretations this exceptional status of space can be seen as well. Using the example of Bohmian Mechanics in the reading of moderate structural realism, the position of a particle remains the only intrinsic property left\footnote{It is remarkable that Bohmian Mechanics follows the position of moderate structural realism insofar as there are objects on the ontological level but with a lack of intrinsic properties.}. All other properties turn out to be relational and support the idea of an abandoning of the object-character of quantum systems. \newline

\subsection{Quambiguity}

In \cite[]{krizek2017ockham}, we presented a quality of Quantum Mechanics that reflects the ambiguity in the ontological commitments that the different interpretations of Quantum Mechanics imply. To reflect this ambiguous character of Quantum Mechanics, we coined the term "Quambiguity". \newline

It has been shown by \cite[]{quine1951main} and others that there is an underdetermination of science in general, but specifically in Quantum Mechanics it seems that this underdetermination is raised to a new level. Quantum Mechanics allows a variety of interpretations that advance completely orthogonal ontologies. Is this "Qambiguity" a feature of the formalism, and therefore a question of the representation of the empirical facts? Or, on the other hand, does "Quambiguity" give an insight into the ontological structure of the world beyond Quantum Mechanics, and what is the conclusion if it is possible to involve such a variety of ontological conceptions with one formalism and one set of empirical facts? \newline

We cannot give an answer on these serious questions, but we can give some conjectures. It seems possible again to take a realist and anti-realist position on the question on the nature and reason for Quambiguity. A anti-realist position could be to assert that the ambiguity in the ontological commitment is an indicator for the redundancy of any ontological commitment. This could be the position of a radical Empiricism, or some related view. A realist position could be to question the nature of ontological entities that we rely on. Usually, realism is related to an object-realism, asserting objects as the entities on the ontological level, which possess unique well-defined properties. This object-realism could be questioned by Quambiguity, as is done in the approaches of structural realism \cite[]{esfeld2008moderate}, \cite[]{Ladyman1998409},\cite[]{worrall1989structural}. \newline

The variety of possible ontologies in the interpretations of Quantum Mechanics gives the impression that the ontological level allows arbitrary ontological commitments. This impression is wrong; in fact Kochen-Specker Theorem , Bell inequalities and others ruled out a whole branch of ontologies that are not realised in nature \cite[]{kochen1975problem}, \cite[]{zeilinger2002bell}. This demonstrates, according to the presented theorem on ontological truth, that there is indeed progress in the quest for ontological truth\cite[]{krizek2017classification}; nonetheless there might be no final answer to it.

\section{Conclusion}

As reality is such an ambiguous conception, we provided an overview on classical aspects of reality and reality in philosophy of science. We provided definitions for Bell realism and realism concepts in Quantum Mechanics. 

Since the formalism of Quantum Mechanics involves highly non-classical concepts like entanglement, the classical conceptions of reality, like object ontology with intrinsic properties, fails to represent an appropriate ontological conception. \newline

We propose that the structural realist approaches in philosophy of science brings a new perspective into the discussion on the interpretation of Quantum Mechanics. Since the object ontology was a highly successful conception and compatible with our common sense experiences, it is deeply anchored in our thinking and our scientific theories. But though it is so well founded in our common sense experience, object ontology was not a very much successful conception in the past at all. All assumed objects, turned out to be constructed out of deeper building blocks and physics never came to a point to really face an object at all. Atomism showed us the road to these fundamental building blocks of matter, and Quantum Mechanics set the rules of their behaviour. And if we ask ourselves, if these fundamental building blocks, these elementary particles, are objects on the conceptual and ontological level, then we have no good reason to believe that they are. \newline 


The formalism of Quantum Mechanics suggests a modification of our philosophical conceptions about reality in several possible ways. One conception that we might have to give up is the ontological conception of objects. Structural realism suggests that what is out there is not an object-like entity that carries properties; it suggests that what is there is a structure, that can be revealed by our best scientific theories. With this approach, structural realism on the one hand complies with the miracle argument of scientific realism, and on the other hand considers the critique of Pessimistic Meta Induction, in the sense that our best theories always speak about structure and it is structure that reflects the ontological elements. By this change of perspective on structures as the primary ontological relevant entities, structural realism resolves the critique of Pessimistic Meta Induction. What is left open by the discussions on structural realism is the relevance of objects. In that respect, the three positions of epistemic, ontic and moderate ontic structural realism differ vastly. Also the interpretations of Quantum Mechanics provide no coherent answer, but Bohmian mechanics for instance emphasizes objects that carry no intrinsic properties anymore, only the position of quantum objects is relevant. In the abandoning of intrinsic properties of objects, Bohmian mechanics agrees with Copenhagen views, except the position of the quantum system. The position plays a distinguished role in Bohmian mechanics, this distinguished role reflects in the formalism of Quantum Mechanics as the position is assigned to a infinite-dimensional Hilbert space. \newline

The De Broglie-Bohm interpretation emphasizes conceptions like active information as something that exists between the entities as a kind of relation or structure in the sense of structural realism, but not in the sense of an subjective information like in QBism. Information in the De Broglie-Bohm interpretation is seen as a formative cause related to the conceptions by Aristotle, and not as an entity related to knowledge or perception. \newline

In QBism the quantum state represents the belief of a specific agent on the behaviour of a quantum system; there the structure of structural realism would be instantiated by the belief of the specific agent. Since the belief of a specific agent is not an objective quantity it turns out that Quantum Mechanics in the QBist reading is a mixture of subjective and objective elements. In the light of the structural realism movement we proposed that QBism suggests to define a new category of structural realism, doxastic structural realism. \newline

Doxastic structural realism is the position of a structural realism neither in the sense of ontic structural realism nor epistemic structural realism, adopting the QBist interpretation of Quantum Mechanics. The position states that there are agents in the world and their best strategy on decisions of how the world behaves is described by Quantum Mechanics. \newline

The structures in doxastic structural realism are on the one hand the subjective beliefs of agents performing Quantum Mechanics, and on the other hand the objective quantum dimension assigned to the quantum systems and their interrelations. In the spirit of QBism it is more a participatory realism than a revelation about what the world really is. There is no world fully independent of the agent. \newline

The Brukner-Zeilinger interpretation emphasizes information as foundational entity. The quantum state describes the knowledge of a hypothetical observer of the specific quantum system. The De Broglie-Bohm interpretation agrees with the Brukner-Zeilinger interpretation in the sense that information is considered as foundational, even if the De Broglie-Bohm interpretation understands it differently than the Brukner-Zeilinger interpretation. QBism and Brukner-Zeilinger interpretation agree on the nature of information as something related to knowledge, perception or belief of observers, although they disagree on the question if this information is objective or subjective. \newline

Specifically Entanglement turns out to be a key feature that demonstrates this structural realism in Quantum Mechanics. Entangled particles do not possess properties in the individuals; what can be assigned instead, are relations between the entangled particles. The entangled quantum state is the representation of these relations.  \newline 

After all, the status of objects in the approaches of structural realism and the interpretation of Quantum Mechanics remains an open question, but its ontological relevance is strongly damaged. The old and very successful conception of objects seems to have come to an end. What is out there on the ontological level is something different. \newline

It remains an open question how it is possible in the framework of structural realism that these structures lead to such a manifold of feasible interpretations that are coherent and overlapping in their formalism and empirical predictions, but orthogonal in their ontological commitments.

\section{Acknowledgments}

I would like to thank the head of the Quantum Particle Workgroup Beatrix Hiesmayr for her support and Basil Hiley, Chris Fuchs, Stefanie Lietze, Emil Simeonov, Agnes Rettelbach, and several other colleagues for fruitful discussions. For proofreading I would like to thank Daniel Moran. For the realisation of the graphical elements and illustrations, I would like to thank Isabella W\"ober and Daniel Alpar.

\bibliography{Quantenmechanik}

\begin{thebibliography}{}

\bibitem[Andreas, 2016]{sep-theoretical-terms-science}
Andreas, H. (2016).
\newblock Theoretical terms in science.
\newblock In Zalta, E.~N., editor, {\em The Stanford Encyclopedia of
  Philosophy}. Metaphysics Research Lab, Stanford University, winter 2016
  edition.

\bibitem[Bell, 1964]{bell1964a}
Bell, J.~S. (1964).
\newblock {On the Einstein-Podolsky-Rosen paradox}.
\newblock {\em Physics}, 1:195--200.

\bibitem[Bell, 1981]{bell1981bertlmann}
Bell, J.~S. (1981).
\newblock Bertlmann's socks and the nature of reality.
\newblock {\em Le Journal de Physique Colloques}, 42(C2):C2--41.

\bibitem[Bohm and Curd, 1981]{bohm1981wholeness}
Bohm, D. and Curd, M. (1981).
\newblock Wholeness and the implicate order.

\bibitem[Bohm and Hiley, 1984]{bohm1984measurement}
Bohm, D. and Hiley, B.~J. (1984).
\newblock Measurement understood through the quantum potential approach.
\newblock {\em Foundations of Physics}, 14(3):255--274.

\bibitem[Bohm et~al., 1987]{bohm1987ontological}
Bohm, D., Hiley, B.~J., and Kaloyerou, P.~N. (1987).
\newblock An ontological basis for the quantum theory.
\newblock {\em Physics Reports}, 144(6):321--375.

\bibitem[Bohr, 1935]{bohr1935can}
Bohr, N. (1935).
\newblock Can quantum-mechanical description of physical reality be considered
  complete?
\newblock {\em Physical review}, 48(8):696.

\bibitem[Braver, 2007]{braver2007thing}
Braver, L. (2007).
\newblock {\em A thing of this world: A history of continental anti-realism}.
\newblock Northwestern University Press.

\bibitem[Brukner, 2017]{brukner2017quantum}
Brukner, {\v{C}}. (2017).
\newblock On the quantum measurement problem.
\newblock In {\em Quantum [Un] Speakables II}, pages 95--117. Springer.

\bibitem[Brukner and Zeilinger, 1999]{brukner1999operationally}
Brukner, {\v{C}}. and Zeilinger, A. (1999).
\newblock Operationally invariant information in quantum measurements.
\newblock {\em Physical Review Letters}, 83(17):3354.

\bibitem[Brukner and Zeilinger, 2003]{brukner2003information}
Brukner, {\v{C}}. and Zeilinger, A. (2003).
\newblock Information and fundamental elements of the structure of quantum
  theory.
\newblock In {\em Time, quantum and information}, pages 323--354. Springer.

\bibitem[Cabello, 2015]{cabello2015interpretations}
Cabello, A. (2015).
\newblock Interpretations of quantum theory: a map of madness.
\newblock {\em arXiv preprint arXiv:1509.04711}.

\bibitem[Chalmers, 2009]{chalmers2009scientist}
Chalmers, A. (2009).
\newblock {\em The scientist's atom and the philosopher's stone: How science
  succeeded and philosophy failed to gain knowledge of atoms}, volume 279.
\newblock Springer.

\bibitem[Chalmers, 2014]{sep-atomism-modern}
Chalmers, A. (2014).
\newblock Atomism from the 17th to the 20th century.
\newblock In Zalta, E.~N., editor, {\em The Stanford Encyclopedia of
  Philosophy}. Metaphysics Research Lab, Stanford University, winter 2014
  edition.

\bibitem[Clauser et~al., 1969]{clauser1969Bellexperiment}
Clauser, J.~F., Horne, M.~A., Shimony, A., and Holt, R.~A. (1969).
\newblock Proposed experiment to test local hidden-variable theories.
\newblock {\em Phys. Rev. Lett.}, 23:880--884.

\bibitem[D{\"u}rr and Teufel, 2009]{durr2009bohmian}
D{\"u}rr, D. and Teufel, S. (2009).
\newblock {\em Bohmian mechanics: the physics and mathematics of quantum
  theory}.
\newblock Springer Science \& Business Media.

\bibitem[Einstein, 1948]{einstein1948quantum}
Einstein, A. (1948).
\newblock Quantum mechanics and reality.
\newblock {\em Dialectica}, 2(3-4):320--324.

\bibitem[Einstein et~al., 1935]{einstein1935can}
Einstein, A., Podolsky, B., and Rosen, N. (1935).
\newblock Can quantum-mechanical description of physical reality be considered
  complete?
\newblock {\em Physical review}, 47(10):777.

\bibitem[Ellis, 2005]{ellis2005physical}
Ellis, B. (2005).
\newblock Physical realism.
\newblock {\em Ratio}, 18(4):371--384.

\bibitem[Englert, 2013]{englert2013quantum}
Englert, B.-G. (2013).
\newblock On quantum theory.
\newblock {\em arXiv preprint arXiv:1308.5290}.

\bibitem[Esfeld, 2004]{esfeld2004quantum}
Esfeld, M. (2004).
\newblock Quantum entanglement and a metaphysics of relations.
\newblock {\em Studies in History and Philosophy of Science Part B: Studies in
  History and Philosophy of Modern Physics}, 35(4):601--617.

\bibitem[Esfeld, 2007]{esfeld2007metaphysics}
Esfeld, M. (2007).
\newblock Metaphysics of science between metaphysics and science.
\newblock {\em Grazer Philosophische Studien}, 74(1):199--213.

\bibitem[Esfeld, 2008]{esfeld2008naturphilosophie}
Esfeld, M. (2008).
\newblock {\em Naturphilosophie als Metaphysik der Natur}.
\newblock Suhrkamp Frankfurt am Main.

\bibitem[Esfeld, 2013]{esfeld2013ontic}
Esfeld, M. (2013).
\newblock Ontic structural realism and the interpretation of quantum mechanics.
\newblock {\em European Journal for Philosophy of Science}, 3(1):19--32.

\bibitem[Esfeld and Lam, 2008]{esfeld2008moderate}
Esfeld, M. and Lam, V. (2008).
\newblock Moderate structural realism about space-time.
\newblock {\em Synthese}, 160(1):27--46.

\bibitem[Feyerabend, 1993]{feyerabend1993against}
Feyerabend, P. (1993).
\newblock {\em Against method}.
\newblock Verso.

\bibitem[Fine, 2009]{fine2009shaky}
Fine, A. (2009).
\newblock {\em The shaky game}.
\newblock University of Chicago Press.

\bibitem[Fuchs, 2017]{fuchs2017notwithstanding}
Fuchs, C.~A. (2017).
\newblock Notwithstanding bohr, the reasons for qbism.
\newblock {\em arXiv preprint arXiv:1705.03483}.

\bibitem[Fuchs and Stacey, 2016]{fuchs2016qbist}
Fuchs, C.~A. and Stacey, B.~C. (2016).
\newblock Qbist quantum mechanics: Quantum theory as a hero's handbook.
\newblock {\em arXiv preprint arXiv:1612.07308}.

\bibitem[Gallie, 1955]{gallie1955essentially}
Gallie, W.~B. (1955).
\newblock Essentially contested concepts.
\newblock In {\em Proceedings of the Aristotelian society}, volume~56, pages
  167--198. JSTOR.

\bibitem[Gardner, 1972]{gardner1972quantum}
Gardner, M.~R. (1972).
\newblock Quantum-theoretical realism: Popper and einstein v. kochen and
  specker.
\newblock {\em The British Journal for the Philosophy of Science},
  23(1):13--23.

\bibitem[Hilary, 1988]{putnam1988representation}
Hilary, P. (1988).
\newblock {\em Representation and reality}.

\bibitem[Holland, 1995]{holland1995quantum}
Holland, P.~R. (1995).
\newblock {\em The quantum theory of motion: an account of the de Broglie-Bohm
  causal interpretation of quantum mechanics}.
\newblock Cambridge university press.

\bibitem[Khlentzos, 2016]{sep-realism-sem-challenge}
Khlentzos, D. (2016).
\newblock Challenges to metaphysical realism.
\newblock In Zalta, E.~N., editor, {\em The Stanford Encyclopedia of
  Philosophy}. Metaphysics Research Lab, Stanford University, winter 2016
  edition.

\bibitem[Khrennikov, 2016]{khrennikov2016reflections}
Khrennikov, A. (2016).
\newblock Reflections on zeilinger--brukner information interpretation of
  quantum mechanics.
\newblock {\em Foundations of Physics}, 46(7):836--844.

\bibitem[Khrennikov et~al., 2012]{khrennikov2012vaxjo}
Khrennikov, A., Khrennikov, A., Atmanspacher, H., Migdall, A., and Polyakov, S.
  (2012).
\newblock V{\"a}xj{\"o} interpretation of wave function: 2012.
\newblock In {\em AIP Conference Proceedings}, volume 1508, pages 244--252.
  AIP.

\bibitem[Kochen and Specker, 1975]{kochen1975problem}
Kochen, S. and Specker, E.~P. (1975).
\newblock The problem of hidden variables in quantum mechanics.
\newblock In {\em The Logico-Algebraic Approach to Quantum Mechanics}, pages
  293--328. Springer.

\bibitem[Kofler and Zeilinger, 2013]{kofler2013quantum}
Kofler, J. and Zeilinger, A. (2013).
\newblock Quantum information and randomness.
\newblock {\em arXiv preprint arXiv:1301.2515}.

\bibitem[Kox, 2014]{kox2014einstein}
Kox, A.~J. (2014).
\newblock Einstein on statistical physics - fluctuations and atomism.
\newblock In {\em The Cambridge companion to Einstein}, pages 103--116.
  Cambridge University Press.

\bibitem[Kragh, 2012]{kragh2012niels}
Kragh, H. (2012).
\newblock {\em Niels Bohr and the quantum atom: The Bohr model of atomic
  structure 1913-1925}.
\newblock OUP Oxford.

\bibitem[Krizek, 2017a]{krizek2017classification}
Krizek, G.~C. (2017a).
\newblock A classification scheme for interpretations of quantum mechanics.
\newblock {\em forthcoming}.

\bibitem[Krizek, 2017b]{krizek2017einstein}
Krizek, G.~C. (2017b).
\newblock Einsteins 1935 papers.
\newblock {\em forthcoming}.

\bibitem[Krizek, 2017c]{krizek2017ockham}
Krizek, G.~C. (2017c).
\newblock Ockham's razor and the interpretations of quantum mechanics.
\newblock {\em arXiv preprint arXiv:1701.06564}.

\bibitem[Kuhn, 1962]{kuhn1962structure}
Kuhn, T.~S. (1962).
\newblock {\em The Structure of Scientific Revolutions}.
\newblock University of Chicago Press.

\bibitem[Ladyman, 1998a]{ladyman1998-LADWIS}
Ladyman, J. (1998a).
\newblock What is structural realism?
\newblock {\em Studies in History and Philosophy of Science Part A},
  29(3):409--424.

\bibitem[Ladyman, 1998b]{Ladyman1998409}
Ladyman, J. (1998b).
\newblock What is structural realism?
\newblock {\em Studies in History and Philosophy of Science Part A}, 29(3):409
  -- 424.

\bibitem[Ladyman, 2016]{sep-structural-realism}
Ladyman, J. (2016).
\newblock Structural realism.
\newblock In Zalta, E.~N., editor, {\em The Stanford Encyclopedia of
  Philosophy}. Metaphysics Research Lab, Stanford University, winter 2016
  edition.

\bibitem[Landsman, 2006]{landsman2006champions}
Landsman, N.~P. (2006).
\newblock When champions meet: Rethinking the bohr--einstein debate.
\newblock {\em Studies in History and Philosophy of Science Part B: Studies in
  History and Philosophy of Modern Physics}, 37(1):212--242.

\bibitem[Laudan, 1981]{laudan1981confutation}
Laudan, L. (1981).
\newblock A confutation of convergent realism.
\newblock {\em Philosophy of science}, 48(1):19--49.

\bibitem[Ludwig, 1978]{ludwig1978grundstrukturen}
Ludwig, G. (1978).
\newblock {\em Die Grundstrukturen einer physikalischen Theorie}.
\newblock Springer-Verlag.

\bibitem[Ludwig and Thurler, 2007]{ludwig2007new}
Ludwig, G. and Thurler, G. (2007).
\newblock {\em A new foundation of physical theories}.
\newblock Springer Science \& Business Media.

\bibitem[Mittelstaedt, 1978]{mittelstaedt1978quantum}
Mittelstaedt, P. (1978).
\newblock Quantum logic.
\newblock In {\em Physical Theory as Logico-Operational Structure}, pages
  153--166. Springer.

\bibitem[Morganti, 2004]{morganti2004preferability}
Morganti, M. (2004).
\newblock On the preferability of epistemic structural realism.
\newblock {\em Synthese}, 142(1):81--107.

\bibitem[Psillos, 2005]{psillos2005scientific}
Psillos, S. (2005).
\newblock Scientific realism and metaphysics.
\newblock {\em Ratio}, 18(4):385--404.

\bibitem[Putnam, 1981]{putnam1981reason}
Putnam, H. (1981).
\newblock {\em Reason, truth and history}, volume~3.
\newblock Cambridge University Press.

\bibitem[Quine, 1951]{quine1951main}
Quine, W.~V. (1951).
\newblock Main trends in recent philosophy: Two dogmas of empiricism.
\newblock {\em The philosophical review}, pages 20--43.

\bibitem[Reichenbach, 1944]{reichenbach1944philosophic}
Reichenbach, H. (1944).
\newblock {\em Philosophic foundations of quantum mechanics}.
\newblock Courier Corporation.

\bibitem[Roecklein, 2015]{roecklein2015locke}
Roecklein, R.~J. (2015).
\newblock {\em Locke, Hume, and the Treacherous Logos of Atomism: The Eclipse
  of Democratic Values in the Early Modern Period}.
\newblock Lexington Books.

\bibitem[Ross, 1928]{ross1928works}
Ross, W.~D. (1928).
\newblock The works of aristotle translated into english.

\bibitem[Ruark, 1935]{ruark1935quantum}
Ruark, A.~E. (1935).
\newblock Is the quantum-mechanical description of physical reality complete?
\newblock {\em Physical Review}, 48(5):466.

\bibitem[Sankey, 2008]{sankey2008scientific}
Sankey, H. (2008).
\newblock {\em Scientific realism and the rationality of science}.
\newblock Ashgate Publishing, Ltd.

\bibitem[Sneed, 1976]{sneed1976philosophical}
Sneed, J.~D. (1976).
\newblock Philosophical problems in the empirical science of science: A formal
  approach.
\newblock {\em Erkenntnis}, 10(2):115--146.

\bibitem[Suppe, 1977]{suppe1977structure}
Suppe, F. (1977).
\newblock {\em The structure of scientific theories}.
\newblock University of Illinois Press.

\bibitem[Tegmark, 2004]{tegmark2004parallel}
Tegmark, M. (2004).
\newblock Parallel universes.
\newblock {\em Science and ultimate reality}, pages 459--491.

\bibitem[Timpson, 2003]{timpson2003applicability}
Timpson, C.~G. (2003).
\newblock The applicability of shannon information in quantum mechanics and
  zeilinger's foundational principle.
\newblock {\em Philosophy of Science}, 70(5):1233--1244.

\bibitem[Timpson, 2013]{timpson2013quantum}
Timpson, C.~G. (2013).
\newblock {\em Quantum information theory and the foundations of quantum
  mechanics}.
\newblock OUP Oxford.

\bibitem[van Inwagen and Sullivan, 2017]{sep-metaphysics}
van Inwagen, P. and Sullivan, M. (2017).
\newblock Metaphysics.
\newblock In Zalta, E.~N., editor, {\em The Stanford Encyclopedia of
  Philosophy}. Metaphysics Research Lab, Stanford University, spring 2017
  edition.

\bibitem[von Meyenn, 2010]{vonMeyenn2010entdeckung}
von Meyenn, K. (2010).
\newblock {\em Eine Entdeckung von ganz au{\ss}erordentlicher Tragweite:
  Schr{\"o}dingers Briefwechsel zur Wellenmechanik und zum Katzenparadoxon}.
\newblock Springer-Verlag.

\bibitem[Von~Neumann, 1955]{von1955mathematical}
Von~Neumann, J. (1955).
\newblock {\em Mathematical foundations of quantum mechanics}.
\newblock Number~2. Princeton university press.

\bibitem[Weihs et~al., 1998]{weihs1998violation}
Weihs, G., Jennewein, T., Simon, C., Weinfurter, H., and Zeilinger, A. (1998).
\newblock Violation of bell's inequality under strict einstein locality
  conditions.
\newblock {\em Physical Review Letters}, 81(23):5039.

\bibitem[Wigner, 1960]{wigner1960unreasonable}
Wigner, E.~P. (1960).
\newblock The unreasonable effectiveness of mathematics in the natural
  sciences. richard courant lecture in mathematical sciences delivered at new
  york university, may 11, 1959.
\newblock {\em Communications on pure and applied mathematics}, 13(1):1--14.

\bibitem[Worrall, 1989]{worrall1989structural}
Worrall, J. (1989).
\newblock Structural realism: the best of both worlds?
\newblock {\em Dialectica}, 43(1-2):99--124.

\bibitem[Zeilinger, 1999]{zeilinger1999foundational}
Zeilinger, A. (1999).
\newblock A foundational principle for quantum mechanics.
\newblock {\em Foundations of Physics}, 29(4):631--643.

\bibitem[Zeilinger, 2002]{zeilinger2002bell}
Zeilinger, A. (2002).
\newblock Bell’s theorem, information and quantum physics.
\newblock In {\em Quantum [Un] speakables}, pages 241--254. Springer.

\end{thebibliography}

\end{document}